\documentstyle[epsf,sprocl]{article}

\def\Journal#1#2#3#4{{#1} {\bf #2}, #3 (#4)}

\def\APJ{\em Astrophysical Journal}
\def\AJ{\em Astronomical Journal}
\def\CPC{\em Comput. Phys. Comm.}
\def\NEWA{\em New Astronomy}
\def\MNRAS{\em MNRAS}

\def\be{\begin{equation}}
\def\ee{\end{equation}}
\def\bea{\begin{eqnarray}}
\def\eea{\end{eqnarray}}

\begin{document}

\title{SIMULATING THE LYMAN ALPHA FOREST}

\author{Marie E. Machacek}

\address{Department of Physics, Northeastern University, 
Boston, MA 02115} 

\author{Greg L. Bryan}

\address{Princeton University Observatory, Princeton, NJ 08544}

\author{Peter Anninos}

\address{Laboratory for Computational Astrophysics, National Center for\\
Supercomputing Applications, 405 Matthews Ave, Urbana, IL 61801}

\author{Avery Meiksin}

\address{Institute of Astronomy, University of Edinburgh\\
Royal Observatory, Blackford Hill, Edinburgh EH9 3HJ, UK} 

\author{Michael L. Norman}

\address{Astronomy Department, University of Illinois at Urbana-Champaign\\
Urbana, IL 61801}


\maketitle\abstracts{In this paper we review the importance of the Lyman 
alpha forest as a probe of structure formation in the universe. We first 
discuss the statistics used to describe the Lyman alpha forest and the 
numerical techniques used to produce simulated spectra of the forest from a 
given cosmological model.  We then discuss the physical picture of the 
absorbing structures that emerges from these numerical simulations. Finally, 
we comment on how two of the statistics, the slope of the column density 
distribution and the $b$ parameter distribution, may be used to constrain 
competing cosmologies. }

\section{Introduction}
Technological advances over the past two decades have resulted in a wealth of
new, good quality observations that allow us for the first time to 
quantitatively as well as qualitatively constrain cosmological models.  
Among this data are high resolution spectra of quasar light~\cite{data} that 
reveal hundreds of absorption features blueward of the Lyman alpha emission 
peak ($\lambda = 121.56701\rm\, nm$ in the rest frame of the quasar's neutral 
hydrogen, but appropriately redshifted due to cosmological expansion).  
Usually a handful of the narrowest features can be identified with heavy 
elements such as iron or carbon.  Occasionally the spectrum contains a broad 
saturated feature, most probably associated with a galactic halo, where the 
integrated number density of neutral hydrogen along the observer's line of 
sight (the HI column density) may exceed $10^{21}\rm\, cm^{-2}$.  However, 
the vast majority of the spectral lines are due to the absorption of quasar 
light by intervening neutral hydrogen of much lower column densities, 
$10^{12}$ to $10^{17}\rm\, cm^{-2}$.  For redshifts $z \ge 3$ 
this ``Lyman alpha forest'' accounts for more than $90 \%$ of all the 
baryons allowed by nucleosynthesis.  Thus these lines map the 
distribution and properties of matter in the universe at a time when 
structure is rapidly forming yet before most of the baryons have collapsed 
into highly nonlinear objects such as galaxies and galaxy clusters. This
gives the Lyman alpha forest a unique place in the study of structure 
formation. The data enjoy other advantages as well.  Since the quasar 
distribution does not pick out a preferred direction in the sky, the quasar 
lines of sight randomly sample the universe.  The data also provide good 
statistics since the spectrum from each quasar line of sight typically 
contains hundreds of Lyman alpha forest lines.  

\section{Statistics of the Forest}
The quantity measured directly by observers is the flux $F$, defined as the 
transmission probability of light as a function of wavelength, or equivalently
velocity, along a given line of sight.  Once the continuum has been 
determined, absorption lines in the resulting spectrum are usually fit using 
Voigt profile functions.  For HI the gaussian component dominates and may be 
characterised by the HI column density ($N_{HI}$) and the Doppler width 
parameter $b$ for the line.  In order to connect the spectrum to the 
underlying properties of the absorbing medium, it is useful to consider the 
optical depth $\tau$ related to $F$ by $F = \exp{(-\tau)}$ where $\tau$ 
depends on the HI density distribution along the line of sight, particle 
velocities and the Lyman alpha absorption cross section.~\cite{huigz}
These quantities can be computed from numerical simulations, as discussed 
below, and thus be used to compare model predictions with observation.        

Statistical measures are needed to characterize the properties of 
the hundreds of observed lines in a given spectrum.  These statistics fall 
into two main classes, those that depend upon the line fitting procedure and 
those that work directly with the raw distribution of flux.  Historically 
most observational data were analysed in terms of the statistics of the Voigt 
profile line fit.  We will focus on two of the most common statistics of this 
kind, $N_{HI}$ and the $b$ parameter distributions.  
The $N_{HI}$ distribution is defined to be the number of lines 
per unit wavelength per column density interval of neutral 
hydrogen. The amplitude of the distribution is not useful to  discriminate 
between models because it scales, up to a weak temperature dependence, as 
$ {( \Omega_b  h^2 )^2/\! \Gamma } $ where $h$ is the dimensionless Hubble 
parameter in units of $100\rm\, km/\!s/\!Mpc$, $\Omega_b$ is the fraction of 
the 
critical density carried in baryons, and $\Gamma$ is the photoionization rate 
at the Lyman alpha edge.  Since $\Gamma$ is not well known, this scale factor
is usually taken as a free parameter determined from the data by either 
fitting the amplitude of the $N_{HI}$ distribution at a particular 
point or computing the average flux decrement at a given redshift.  The 
slope of the $N_{HI}$ distribution is, however, a robust statistic that can 
be used to test models.  The data are well fit by a power law  with 
slope $\beta = -1.5$ for column densities below $10^{14}\rm\, cm^{-2}$, but 
the distribution steepens for higher column densities.  A second statistic, 
the $b$ parameter distribution, counts the number of lines per unit redshift 
of a given Doppler width.  These distributions are more challenging to 
interpret.  This is due, in part, to the fact that the fit itself is not 
unique.  Many of the absorption features found in the data are blends 
of two or more lines. Furthermore the connection between the $b$ 
distribution and the properties of the medium is complex. Absorber 
temperature, gas velocity (Hubble flow and physical) and density 
profile all contribute to the line width.  Because of these difficulties 
new statistics based directly on the raw flux distribution have been 
proposed.~\cite{moments}  These include the flux probability distribution 
and moments of the n-point flux distribution.  Although these statistics 
have the advantage that they are independent of any line fitting program, 
their major disadvantage is that little data have yet to be analysed in this 
way.  We discuss these line independent statistics in more detail 
elsewhere.~\cite{bman,mach} In either case numerical simulations of 
synthetic spectra and their subsequent analysis provide the link between the 
observed spectrum and the underlying cosmological model.

\section{Simulation Strategy}
We consider a suite of the currently most promising hierarchical cosmological
models characterized by the dimensionless Hubble parameter $h$, the fraction
$\Omega$ of the critical density carried in each model component (baryonic 
gas, cold and/or hot dark matter, cosmological constant), and the slope $n$
of the initial power spectrum of density fluctuations. The models~\cite{mach}
 include 
the standard cold dark matter model (SCDM), a flat cold dark matter model 
with nonvanishing cosmological constant (LCDM), a low density cold dark 
matter model (OCDM), a flat cold dark matter model with a tilted power 
spectrum (TCDM), and a critical model with both cold and hot (two massive 
neutrinos) dark matter components (CHDM).  The SCDM, OCDM, and LCDM model 
parameters agree with earlier simulations by Zhang,{\it et al}.~\cite{zhang} 
TCDM and CHDM model parameters are taken from the Grand Challenge Cosmology 
Consortium model comparison project. The initial fluctuation spectra, 
assumed to be gaussian, are normalized using the observed distribution of 
galaxies, although all but SCDM are also consistent with measurements of the 
cosmic microwave background. 

We use a single grid Eulerian code (KRONOS) with a particle-mesh gravity 
solver and a modified piecewise parabolic method to simulate the gas 
hydrodynamics to evolve the system nonlinearly to moderate redshifts 
$z \simeq 2$.~\cite{kronos} The radiation field, assumed spatially constant,
is computed from the observed quasar distribution~\cite{haardt} with 
spectral index $\alpha=1.5$ which reionizes the universe around $z\sim 6$ 
and peaks at $z\simeq 2$. Since nonequilibrium effects can be important, 
we calculate radiative processes such as photoionization, recombination and 
Compton cooling for six chemical species (HI, HII, HeI, HeII, HeIII, e) using 
the method of Anninos, {\it et al}.~\cite{pete} However, we do not include 
radiative transfer, self-shielding, star formation, or feedback from star  
formation and so can not address the physics within the highest density clumps 
($N_{HI}\ge 10^{16}\rm\,cm^{-2}$). We use $256^3$ grid cells in a simulation 
box of length $9.6\rm\, Mpc$ comoving with the universal expansion and 
follow the evolution of $128^3$ ($256^3$) dark matter (gas) particles, 
respectively.  This results in a spatial resolution of $37.5\rm\, kpc$ and 
mass resolutions of 
$2.9\times 10^7\rm\, M_\odot$ ($3.6\times 10^6\rm\, M_\odot$) for the 
dark matter (gas) particles.

The output from numerical simulations of cosmological models are three- 
dimensional snapshots of the state of the universe as the universe ages 
(redshift $z$ decreases).  For each $z$ chosen the simulation returns the 
positions and velocities for each particle, the temperature, 
pressure, and gas density, and the distribution of each chemical species 
throughout the box. By taking slices through the simulation box we can 
directly compare the physical properties of matter in the box. 
In particular the 
distribution of neutral hydrogen traces well the baryon gas distribution even
though by $z\simeq 5$ the universe is almost completely reionized with the 
fraction of gas remaining neutral $\le 10^{-4}$. Furthermore the gas 
distribution traces well the cold dark matter distribution so that probing 
neutral hydrogen by use of Lyman alpha absorption is indeed mapping the 
matter distribution in the universe. In order
to generate synthetic quasar absorption spectra from the simulation data, 
we place a quasar within the box that is well removed from the box edges.  
We then trace the absorption of Lyman alpha 
radiation from the quasar by the neutral hydrogen in the box along $300$ 
randomly generated lines of sight.  The spectra, i.e. transmitted 
flux as a function of velocity or wavelength, so produced can be analysed
in the same way as the observational data, and the statistics, after 
averaging over the $300$ lines of sight, compared with observation.

There is a tension in simulations of the Lyman alpha forest between the need 
for large simulation box size to include sufficient large scale power from the
density fluctuations and the need for high spatial resolution to adequately 
model the absorbing structures. We use SCDM as a typical hierarchical model 
to systematically study these numerical uncertainties.~\cite{bman} 
By varying the simulation box length by factors of two from $2.4\rm\, Mpc$ to 
$19.2\rm\, Mpc$ at $75\rm\, kpc$ fixed spatial resolution, we find that 
the physical properties and statistical measures of the gas discussed 
above are nearly identical for the largest two box sizes demonstrating 
convergence by $9.6\rm\, Mpc$. The major difference due to the presence of 
larger scale modes in the bigger boxes is an increase in the amount of high 
temperature gas due to shock heating around the edges of the more dense 
sheets, filaments and knots. This has little effect on the low density 
regions responsible for the forest.  Similarly we study the effects of 
spatial resolution by fixing the box size at $2.4\rm\, Mpc$ (small so that 
we could obtain high grid resolution) and decreasing the grid spacing by 
factors of two from $75\rm\, kpc$ to $18.75\rm\, kpc$.  
The physical properties of the gas are again quite similar. However there is 
a slight tendency for the higher resolutions to have more low density gas due 
to more small scale power and a better ability to resolve small scale 
features.  Since lower density regions are cooler due to expansion cooling 
and reduced photo-ionization, higher grid resolution also systematically 
produces more low-temperature gas. Again these differences have little effect 
on the column density distributions for neutral hydrogen showing the robust 
nature of this statistic.  On the other hand, the $b$ parameter 
distribution systematically decreases with increased resolution, although its 
shape remains largely invariant.  Similar resolution dependence appears 
in fit independent statistics such as the first moment of the 2-point flux 
distribution.  Thus this effect is not an artifact of the line fitting 
procedure.  We argue instead that the principle cause of this resolution 
dependence is numerical thickening of the density 
profile for low spatial resolution.  The $b$ parameter and fit
independent statistics do appear to be converging at our highest spatial 
resolutions. However, it is clear that a spatial resolution of 
$37.5\rm\, kpc$ is just barely sufficient to model these distributions. 

\section{Physical Picture}
\begin{figure}
\epsfxsize=3.5in
\centerline{\epsfbox{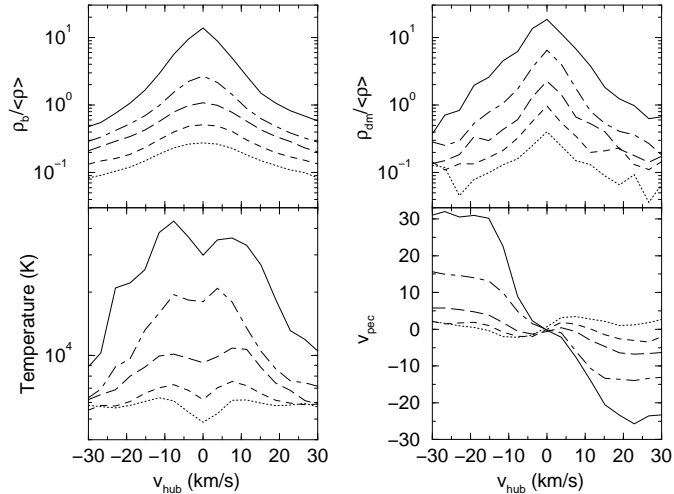}}
\caption{
Mean profiles at $z=3$ for baryon overdensity, dark matter overdensity, gas 
temperature and peculiar velocity distributions projected along the line of 
sight for (bottom to top) $\log(N_{HI}) = 12,12.5,13,13.5,14.5$.
}
\label{fig:prop}
\end{figure}

A consistent picture of the Lyman alpha forest in cold dark matter dominated
cosmologies is now emerging from numerical simulations~\cite{zhang,lars} 
using a variety of numerical techniques and from more analytical 
approximations.~\cite{anal,huigz} In this picture the absorbers that
produce lines with low column densities $N_{HI} < 10^{15}\rm\, cm^{-2}$ at 
$z\approx 3$ are large, unvirialized objects of size $\sim 100\rm\, kpc$ and 
low densities comparable to the cosmic mean~\cite{zhang98} that have
grown by gravitational amplification from primordial density fluctuations.  
In order to refine this picture we select those cells from the SCDM 
simulations~\cite{bman} that correspond to a line in a given column 
density range and compute the mean characteristic (dark matter $\delta_{dm}$ 
and gas $\delta_b$ overdensities, temperature, and peculiar velocity) 
profiles across these structures.  These are plotted for $z=3$ in 
Figure~\ref{fig:prop}. From the overdensity profiles we see that these 
structures range from underdense structures ($\delta \sim 0.2$) 
for $ N_{HI} = 10^{12}$ to mildly nonlinear structures ($\delta \approx 10$) 
for $N_{HI}=10^{14.5}$. In all cases the wings of the profile drop below 
the cosmic mean.  This means that most low column density lines are produced 
by absorbers located in voids.  The velocity plot indicates that these 
absorbers are infalling in comoving coordinates.  However, the infall 
velocities for absorbers with $N_{HI} \le 10^{13.5}$ in this model are less 
than the Hubble expansion so that these lowest column density structures are 
actually expanding in absolute coordinates.   Once the column density 
exceeds $10^{14}\rm\, cm^{-2}$, the qualitative character of the absorber 
changes. The infall velocity becomes greater than the Hubble flow so that 
the object has collapsed.

A careful study of the central regions of these profiles also reveals 
interesting physics.  The dark matter and gas density profiles are remarkably 
similar, except near the peak where the gas exhibits a rounded rather than 
cuspy shape.  When the gas was heated at reionization to roughly 
$15,000\rm\, K$, gas in the center of the peaks found itself 
overpressured and started to expand.  This expansion is seen as a kink 
with positive slope at the center of the velocity profiles for the lower 
column densities and causes the smoothing of the peak of the gas density 
distribution relative to that for dark matter.  The central gas cooled as it 
expanded causing a corresponding dip at the center of the temperature 
profile.  As the column density increases, pressure is less effective 
against gravity in causing the central gas to expand. Thus the kink in the 
velocity profile flattens, the dip in the temperature profile is diminished 
and the peak in the density profile becomes more pronounced.  At the highest 
column densities, $N_{HI} \sim 10^{14}\rm\, cm^{-2}$, when the absorber 
changes from an uncollapsed to a collapsed system, two shocks form on either 
side of the absorber midplane and propagate outwards. This can be seen in 
the velocity distribution at $\pm 10\rm\, km/\!s$ and in the twin 
shock-heated peaks in the temperature distribution. It is intriguing to note 
that this qualitative change in the character of the absorber occurs near 
the column density where carbon abundances are observed to increase 
dramatically~\cite{lu98} and where the column density distribution appears to 
steepen.

\section{Testing Cosmological Models}
\begin{figure}
\epsfxsize=3.5in
\centerline{\epsfbox{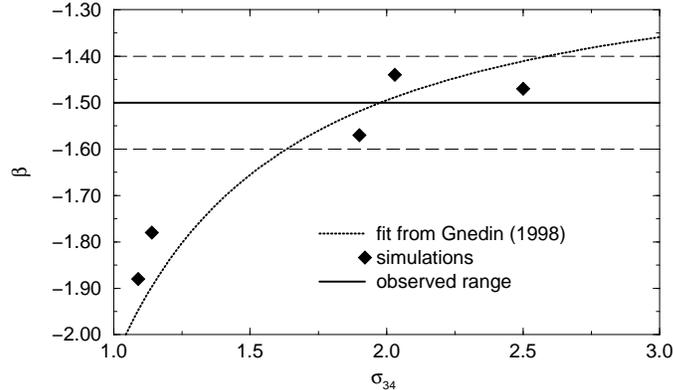}}
\caption{Slope of the column density distribution at $z=3$ as a function of 
$\sigma_{34}$.  Simulation models are from left to right: TCDM, CHDM, SCDM,
LCDM, OCDM 
}
\label{fig:slope}
\end{figure}

In order to use the Lyman alpha forest to discriminate between competing 
cosmologies we need to use our physical understanding of the nature of the 
absorbers to link the measured statistics to the characteristic properties 
that define the model. The column density distribution is an integrated 
quantity relatively insensitive to the detailed shape of the 
HI density profile or spatial resolution of the simulation.  The slope of the 
column density distribution is primarily determined by the power carried in 
fluctuations on small $\sim 200$ -- $500\rm\, kpc$ scales with less power 
producing steeper slopes.  One measure of this small scale power, 
$\sigma_{34}$, is the {\it rms} density fluctuation computed from the linearly
evolved power spectrum at the redshift of interest filtered by a scale 
roughly half of the Jean's length.~\cite{gned98}  In Figure~\ref{fig:slope}
we plot the slope of the $N_{HI}$ distribution versus $\sigma_{34}$ for 
$z=3$. As this figure shows, agreement with observations is found for several 
popular models and favors those models with more power at smaller scales.
\begin{figure}
\epsfxsize=3.5in
\centerline{\epsfbox{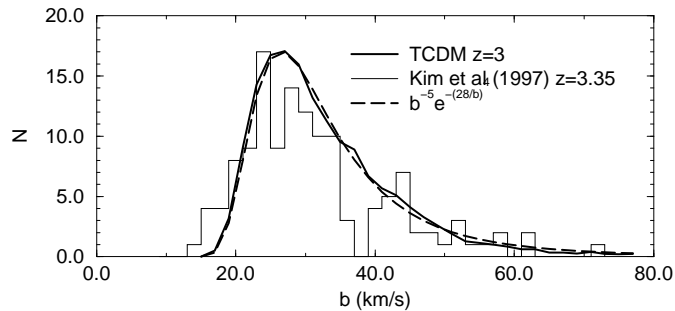}}
\caption{
$b$ parameter distributions for observational data at 
mean $z=3.35$, TCDM simulation at redshift $z=3$, and the analytical profile 
(dashed) $b^{-5}\exp(-b^4_\sigma/\!b^4)$ with $b_\sigma = 28$.
}
\label{fig:b}
\end{figure}

The $b$ parameter distribution at $z=3$ is shown in Figure~\ref{fig:b}. The
shape of this distribution is well reproduced by hierarchical 
cosmologies and follows closely the analytic profile, 
$b^{-5}\exp(-b^4_\sigma/\!b^4)$, of Hui \& Rutledge~\cite{rut} based on 
the gaussian nature of the underlying density and velocity fields.  However,
for our choices of SCDM, LCDM, and OCDM models, the predicted medians 
$b_\sigma$ are substantially too low.~\cite{mach} In order to search for a 
remedy, we need to understand which of the competing physical processes sets 
the scale of the $b$ parameter for low column density absorbers. 
Returning to Figure~\ref{fig:prop} we see that the gas temperatures are too 
low for thermal Doppler broadening to dominate any but the narrowest of 
lines. Similarly the velocities for the gas are small and an increase in the
physical infall velocity would serve to narrow, rather than broaden, the line.
Thus the dominating contributor to the width of the forest lines is the 
shape of the density distribution itself. This shape could be changed by 
increased thermal pressure within the gas, but such a large temperature 
increase seems difficult to obtain.~\cite{bman} Instead we may search 
for key changes in the density distribution affecting the $b$ distribution
due to the differing power spectra of competing cosmologies. Thus   
the Lyman alpha forest holds tremendous promise as a sensitive probe of 
structure formation as our understanding of the connections between the 
spectra and the underlying absorbers continues to improve.

\section*{Acknowledgments}
This work is done under the auspices of the Grand Challenge Cosmology 
Consortium and supported in part by NSF grant ASC-9318185 and NASA 
Astrophysics Theory Program grant NAG5-3923.

\end{document}